\documentclass[journal]{IEEEtran}
\usepackage{stfloats}
\usepackage{amsmath,amsfonts,amssymb,amscd,amsthm}
\usepackage{graphicx,epsfig}
\usepackage{cases}
\usepackage{lscape}
\usepackage{verbatim}
\usepackage{epstopdf}
\usepackage{array}
\usepackage{balance}
\usepackage{multirow}
\usepackage[noadjust]{cite}
\usepackage{bbold}
\usepackage{multirow}
\usepackage{url}
\usepackage{arydshln}
\usepackage{tikz}
\usepackage[normalem]{ulem}
\usepackage{capt-of}
\usepackage{color}
\allowdisplaybreaks[4]

 \definecolor{mylightred}{RGB}{255, 200, 200}
 \definecolor{mylightblue}{RGB}{172, 188, 63}
 \definecolor{mylightgreen}{RGB}{150, 220, 150}

\theoremstyle{definition}
\usepackage{accents}

\makeatletter
\def\maketag@@@#1{\hbox{\m@th\normalfont\normalsize#1}}
\makeatother

\begin{document}

\title{Analysis of November 21, 2021, Kaua`i Island Power System 18--20 Hz Oscillations}

\author{Shuan Dong,~\IEEEmembership{Member,~IEEE},
    Bin Wang,~\IEEEmembership{Senior Member,~IEEE}, Jin Tan,~\IEEEmembership{Senior Member,~IEEE}, \\ Cameron J. Kruse, Brad W. Rockwell,  
    and Anderson Hoke,~\IEEEmembership{Senior Member,~IEEE}
\vspace{-0.5cm}
}

% make the title area
\maketitle

%%%%%%%%%%%%%%%%%%%%%%%%%%%%%%%%%%%%%%%%%%%%%%%%%%%%%%%%%%%%%%%%%%%%%%%%%%%%%%
\begin{abstract} 
This letter discusses the 18--20 Hz oscillation event at~05:30 am on November 21, 2021, in Kaua`i's power system following the trip of an oil power plant. As far as the authors are aware, this is the first report of a transmission system-wide subsynchronous oscillation driven by inverter-based resources (though the system in question is relatively small).  In this letter, we leverage two data-based methods---the dissipating energy flow method and the sub/super-synchronous power flow method---to locate the sources of the oscillation. Also, we build an electromagnetic transient model of the Kaua`i power system and replay the 18-20 Hz oscillation. Finally, we propose \textcolor{black}{two} mitigation methods and validate their effectiveness via numerical simulation.

\end{abstract}


\IEEEpeerreviewmaketitle

\section{Introduction}

\IEEEPARstart{T}{he} accelerating transition toward renewable energy challenges the stability of our power systems. Renewables, such as wind and solar, are inverter-based resources (IBRs) and bring different dynamics and faster interactions to power systems compared to conventional synchronous generators (SGs)~\cite{hatziargyriou2020definition}. For example, as summarized in~\cite{cheng2022real}, IBRs have contributed to numerous subsynchronous oscillation events around the world. Most real-world IBR-related oscillations are local events in large-scale power grids instead of system-wide events, as analyzed here. 

To address IBR-related oscillations, comprehensive post-event analysis is needed to locate oscillation sources, identify fundamental causes, and suggest mitigation methods.  
Most reported IBR-related oscillations are caused by either unwanted interactions between IBRs and system components, e.g., series compensation, or weak (low short-circuit ratio, SCR) interconnections~\cite{cheng2022real}. However, 
detailed case-by-case analysis is still needed for newly observed IBR-related oscillations.
Identifying the mechanisms of IBR-related oscillations is challenging because the grid might include heterogeneous IBR controller designs from different vendors, and also, the well-validated IBR models are typically protected by various non-disclosure agreements or are not available. 

This letter reports a system-wide IBR-related 18--20 Hz oscillation event on Kaua`i's 57 kV transmission system on November 21, 2021. Kaua`i has an isolated power system with no series compensation devices.
Then we analyze this event systematically with both measurement- and model-based methods. Specifically, we first locate the oscillation sources with two measurement-based methods, i.e., dissipating energy flow (DEF)~\cite{chen2012energy,maslennikov2017dissipating} and sub/super-synchronous power flow~\cite{xie2019identifying}. Next, we build an electromagnetic transient (EMT) model of the Kaua`i power system to recreate the initial stage of observed oscillations. Last, by performing parameter sensitivity studies with our EMT model, we propose \textcolor{black}{two} methods to mitigate the oscillations. The effectiveness of our methods is validated through EMT simulations.

\section{Overview and Analysis of Oscillation Event} 
This section first overviews the 18--20 Hz oscillation event on Kaua`i Island. Then, we analyze this event via three steps: locating the oscillation sources, recreating the oscillations in EMT simulations, and proposing \textcolor{black}{two} mitigation methods.

\begin{figure}
\begin{minipage}{0.50\columnwidth}
\centering
\includegraphics[width=1\linewidth]{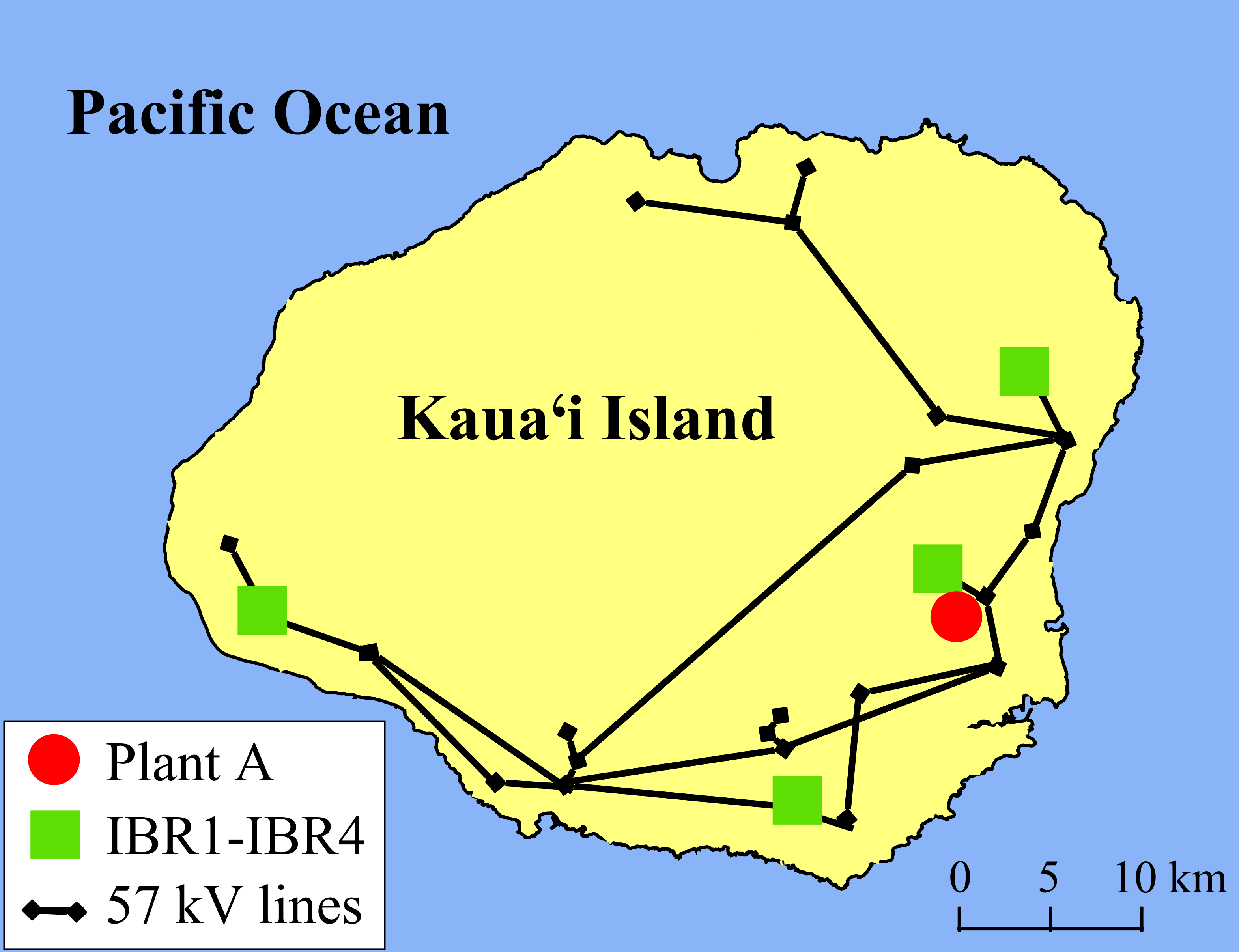} 
\label{fig:redundant}
\end{minipage}
\begin{minipage}{0.45\columnwidth}
\footnotesize
\centering

\vspace{-15pt}
\captionof{table}{\label{tab:kiuc_gen} KIUC Generation Mix Before and After Event}

\vspace{-7.5pt}
\begin{tabular}[b]{c|c|c} 
Time & $t=0^-\,\mathrm{s}$ & $t=60~\mathrm{s}$ \\ \hline
Plant~A & $60.6\%$ & $\,~0.0\%\downarrow$ \\ 
IBR1 & $~4.1\%$ & $14.0\%\uparrow$ \\ 
IBR2 & $~4.6\%$ & $21.0\%\uparrow$ \\ 
IBR3 & $~0.0\%$ & $14.0\%\uparrow$ \\ 
IBR4 & $~4.1\%$ & $23.0\%\uparrow$  \\ 
%Others & $26.7\%$ & $27.0\%$
Biomass & $13.7\%$ & $14.0\%\uparrow$ \\
Hydros & $13.0\%$ & $13.0\%\,-$
\end{tabular}
\end{minipage}

\caption{\label{fig:Kauai_map} Kaua`i map with plant~A and IBR1--IBR4 highlighted.} 
\end{figure}
%%%%%%%%%%%%%%%%%%%%%%%%%%%%%%%%%%%%%%%%%%%%%%

\begin{figure}[t!]
  \centering
  \includegraphics[width=1\linewidth]{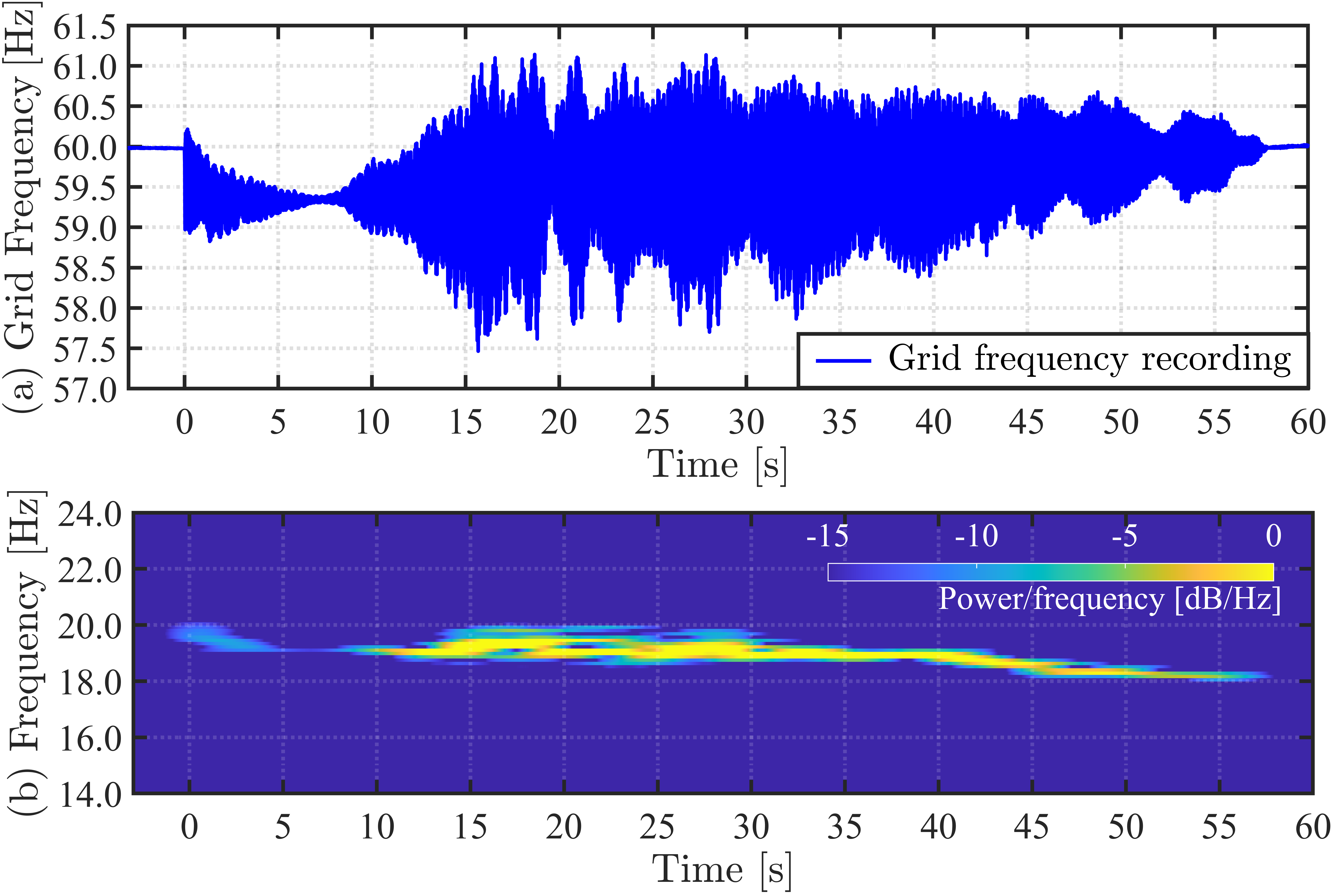}
  
  \caption{Kaua`i frequency recording (starting at 05:30:47.65 am HST on November 21, 2021). (1)~Fundamental frequency. (2)~Spectrogram of frequency showing 18--20 Hz oscillations.} \label{fig:PMRF_Freq}
\end{figure}

\begin{figure}[t!]
  \centering
  \includegraphics[width=1\linewidth]{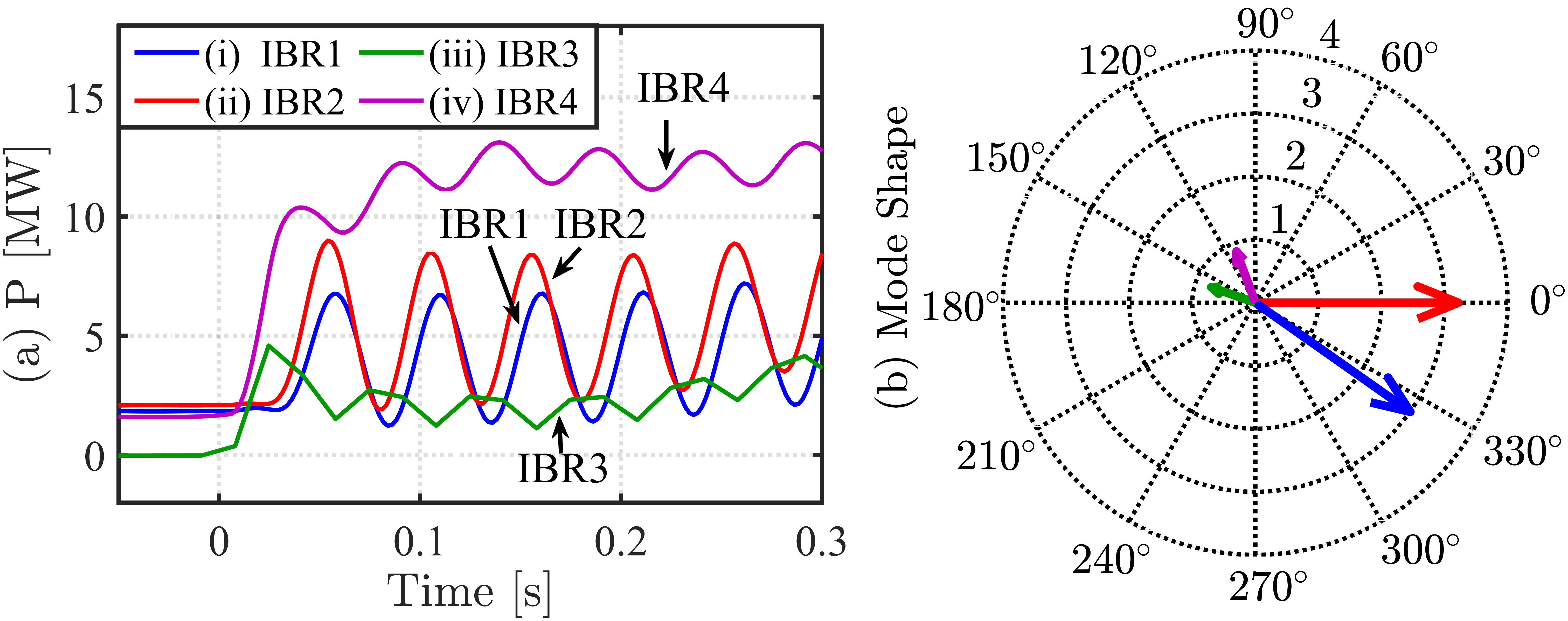}
  \caption{Recorded IBR active-power responses and mode shape of IBRs' dominant 18--20 Hz mode. We can find that IBR1 and IBR2 oscillate against IBR3 and IBR4 after~$t=0~\mathrm{s}$. (a)~Recorded active-power waveforms of IBR1--IBR4. (b)~Mode shape of IBRs' 18--20 Hz mode.} \label{fig:P_and_mode_shape_dfr_v2}
\end{figure}

\subsection{Overview of Kaua`i Island 18--20 Hz Oscillations}

Kaua`i Island, as shown in Fig.~\ref{fig:Kauai_map}, is Hawaii's 4th largest island, with a population of more than 73 thousand. It has an isolated, meshed power system operated by Kaua`i Island Utility Cooperative~(KIUC). In 2021, the peak demand of Kaua`i Island reached~$75.17~\mathrm{MW}$, and~$69.5\%$ of its annual generation was renewable, including solar, hydro, and biomass~\cite{kiuc2021report}.

At 5:30:47 am Hawaii Standard Time (HST) on November 21, 2021 (referred to as~$t=0~\mathrm{s}$ in this letter), an oil plant A that supplied $60.6\%$ of the system load tripped. This represented a very severe $N\!-\!1$ contingency. \textcolor{black}{Figures~\ref{fig:PMRF_Freq}(a) and~\ref{fig:PMRF_Freq}(b), respectively, show the system frequency and frequency spectrogram for the KIUC system.}
\textcolor{black}{We find that} apart from the frequency drop, \textcolor{black}{the plant~A trip} excited 18--20 Hz frequency oscillations that sustained for approximately~$60~\mathrm{s}$ (see spectrogram in Fig.~\ref{fig:PMRF_Freq}(b)). As shown in Table~\ref{tab:kiuc_gen} and Fig.~\ref{fig:P_and_mode_shape_dfr_v2}(a), the inverter-interfaced batteries at four hybrid photovoltaic (PV) power plants, denoted as IBR1--IBR4, provided a fast frequency response within~$50~\mathrm{ms}$. In so doing, IBR1--IBR4 contained the frequency decline within~$1.5~\mathrm{s}$ (see~Fig.~\ref{fig:PMRF_Freq}(a)) and brought the frequency back to~$60~\mathrm{Hz}$ within 1 minute following automatic generation control signals, a remarkable survival given the very large loss of generation. Note that IBR1--IBR3 use grid-following (GFL) control, and IBR4 uses virtual synchronous machine (VSM) control~\cite{dong2016adjusting}. 
Following the plant~A trip, the SCRs at IBR1--IBR4 reduced from 4.3, 3.4, 6.2, and 4.4 to 3.4, 2.6, 5.7, and 2.9, respectively.
Based on the active-power responses of IBR1--IBR4 after~$t=0~\mathrm{s}$, as shown in Fig.~\ref{fig:P_and_mode_shape_dfr_v2}(a), we perform Prony analysis \cite{hauer1990initial} and plot their 18--20 Hz mode shape in Fig.~\ref{fig:P_and_mode_shape_dfr_v2}(b). We find that IBR1 and IBR2 have similar phases and larger oscillation magnitudes, and they oscillate against IBR3 and IBR4 after~$t=0~\mathrm{s}$. 

Although the system regained its stability after 1 minute at the cost of shedding only~$3.7\%$ of its load, the observed 18--20 Hz oscillations shown in Fig.~\ref{fig:PMRF_Freq} pose a serious challenge to the stable operation of the power system. Such oscillations can result in poor power quality, damage power system components, or cause power outages.

\subsection{Analysis of Kaua`i Island 18--20 Hz Oscillations}

In order to address the 18--20 Hz oscillations, we adopt both measurement- and model-based methods and analyze the event with three steps, as follows.

\subsubsection{Oscillation Source Identification} 
\begin{figure}[t!]
  \centering
  \includegraphics[width=1\linewidth]{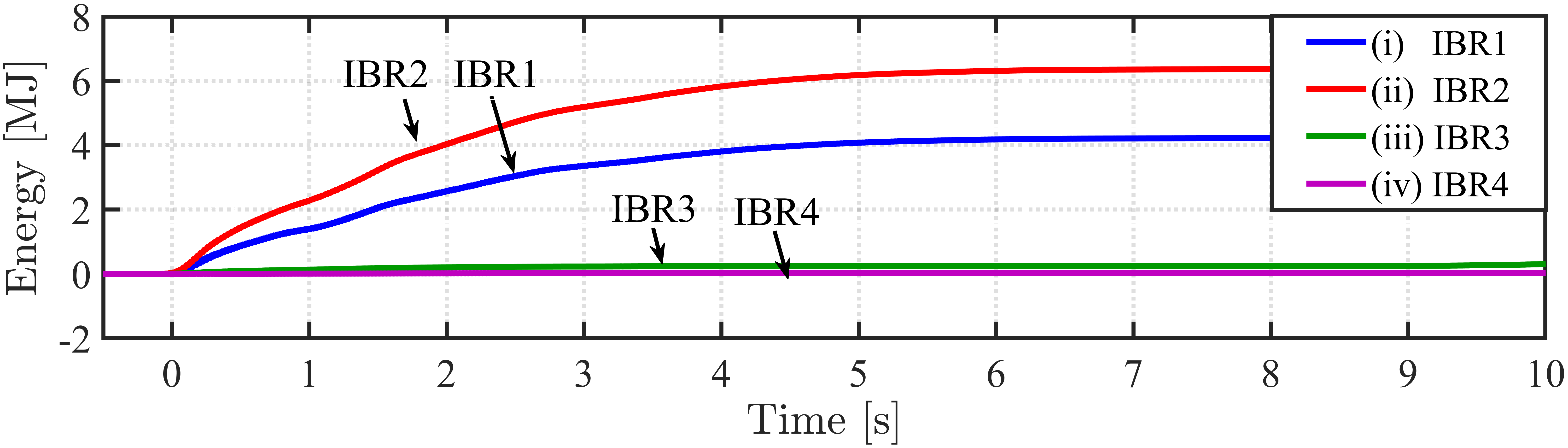}
  \caption{Identification of oscillation sources with the DEF method~\cite{chen2012energy,maslennikov2017dissipating}. This method shows that IBR1 and IBR2 are oscillation sources because they inject oscillation-frequency energy into the grid after~$t=0~\mathrm{s}$.} \label{fig:def}
\end{figure}

\begin{figure}[t!]
  \centering
  \includegraphics[width=1\linewidth]{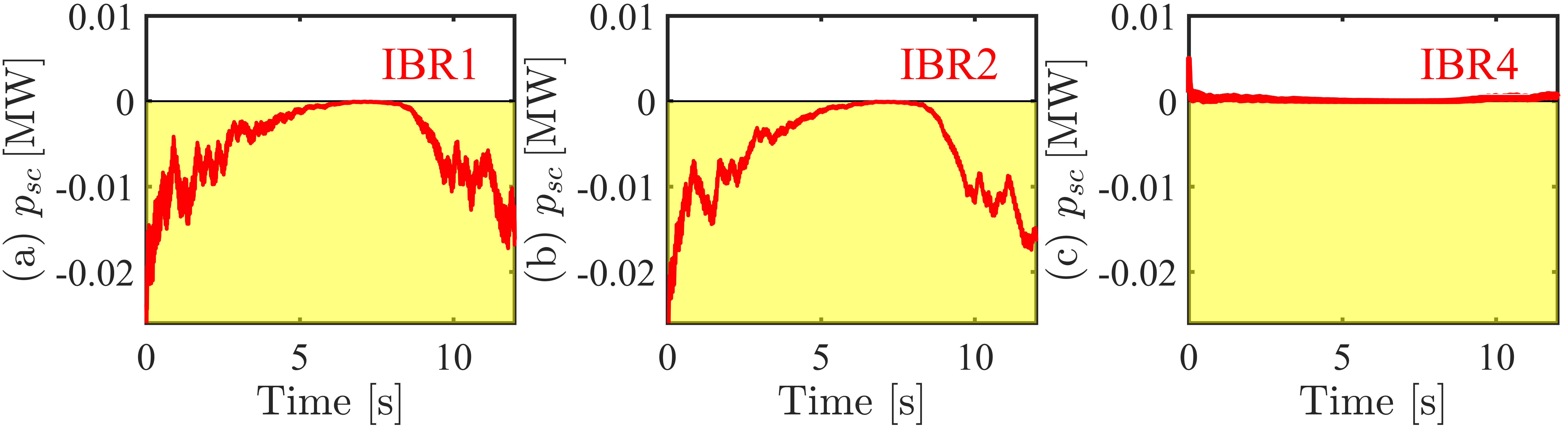}
  \caption{Identification of oscillation sources with sub/super-synchronous~active power~\cite{xie2019identifying}. This method shows that IBR1 and IBR2 are oscillation sources~because they absorb negative sub/super-synchronous active power after~$t=0~\mathrm{s}$.} \label{fig:P_sub_sup_dfr}
\end{figure}

This step leverages two measurement-based methods, i.e., DEF~\cite{chen2012energy,maslennikov2017dissipating} and sub/super-synchronous power flows~\cite{xie2019identifying}, to identify the sources of the 18--20 Hz oscillations. Our inputs for these two methods are, respectively, 60 or~480 Hz phasor data and 15~kHz point-on-wave (PoW) data captured by digital fault recorders (DFRs). The DEF method locates the oscillation sources based on the computed transient energy flow directions. As shown in Fig.~\ref{fig:def}, IBR1 and IBR2 inject large amounts of dissipating energy, whereas IBR3 and IBR4 exchange little energy. Thus, IBR1 and IBR2 are identified as oscillation sources. To crosscheck the analysis result, we also infer the oscillation sources by computing the sub/super-synchronous power flows that correspond to the oscillation frequency. As shown in Fig.~\ref{fig:P_sub_sup_dfr}, only IBR1 and IBR2 absorb negative sub/super-synchronous active power, i.e., $p_{sc}<0$, and are oscillation sources. Note that IBR3 is not analyzed in Fig.~\ref{fig:P_sub_sup_dfr} because its PoW data were not recorded by the DFR. In sum, both methods identify IBR1 and IBR2 as sources of the 18--20 Hz oscillations.

\subsubsection{Oscillation Recreation with EMT Simulations} \label{subsec:recreation}
Our second step is to build the EMT model for Kaua`i's power system and recreate the 18--20 Hz oscillations. First, based on the pre-event power flow recording in KIUC's supervisory control and data acquisition (SCADA) software, we revise the settings of the breakers, generation units, loads, and switched shunts in KIUC's PSS/E model. Specifically, we set all PV outputs to zero because the oscillations appear in early morning with little solar irradiance. Then, we generate the PSCAD network model based on revised PSS/E data with the PRSIM software. Next, we add detailed dynamical models for all online generation units. Among IBR1--IBR4, the black-box EMT models of IBR1 and IBR2 were provided, so we connect them to our PSCAD model. For IBR3 and IBR4, we represent them with generic GFL and VSM models (see~\cite{dong2016adjusting}) and tune their parameters based on the recorded dynamics.

\begin{figure}[t!]
  \centering
  \includegraphics[width=1\linewidth]{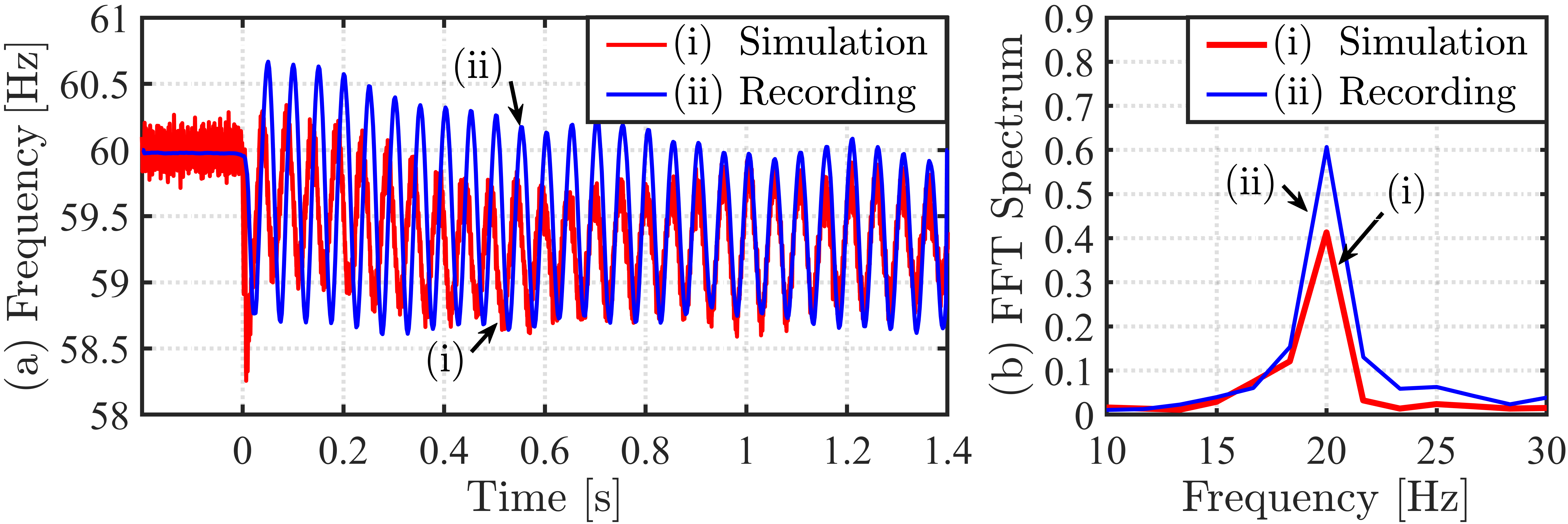}
  \caption{Simulated and recorded grid frequencies measured at IBR1 have similar time-domain responses and FFT spectra, which can validate the EMT model accuracy. (a)~Simulated and recorded grid frequency waveforms. (b)~FFT analysis results of the simulated and recorded grid frequencies.} \label{fig:PMRF_freq_simu}
\end{figure}

\begin{figure}[t!]
  \centering
  \includegraphics[width=1\linewidth]{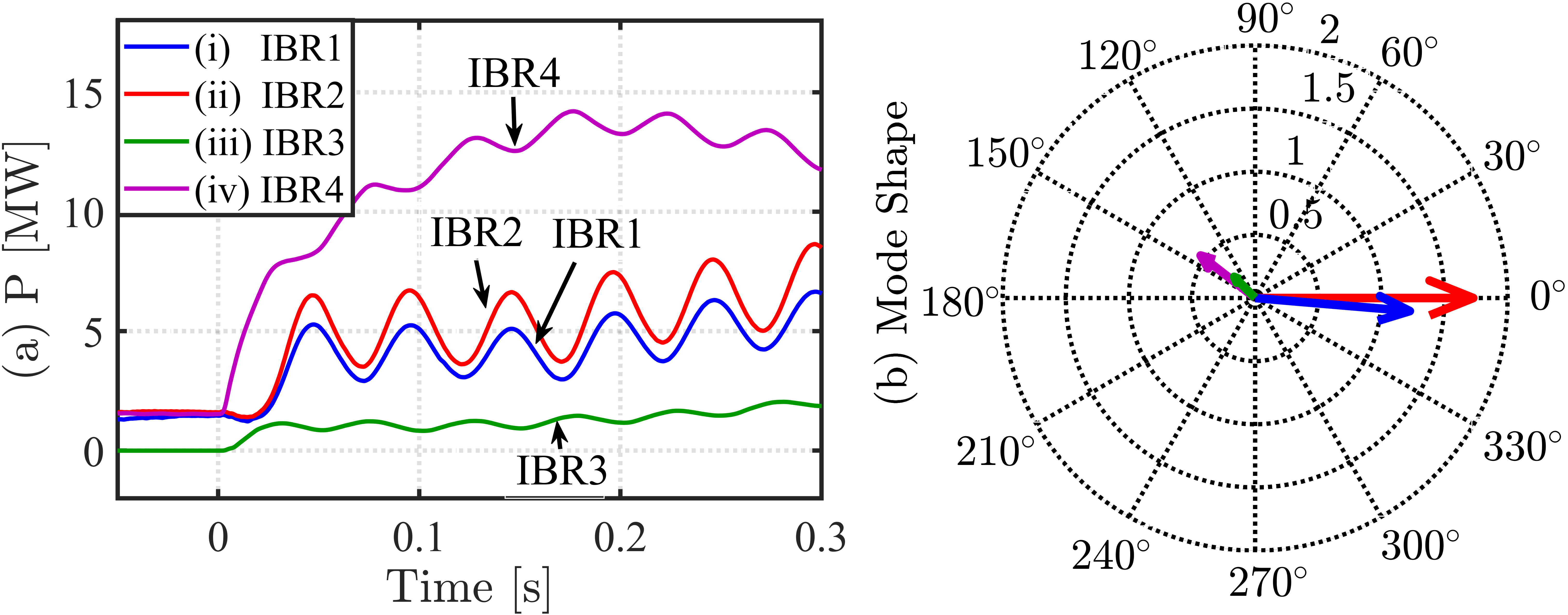}
  \caption{Simulated IBR active-power responses and mode shape of IBRs' dominant 18--20 Hz mode. IBR1 and IBR2 oscillate against IBR3 and IBR4 after~$t=0~\mathrm{s}$. This observation matches the recordings in Fig.~\ref{fig:P_and_mode_shape_dfr_v2} and further validates the accuracy of our EMT model. (a)~Simulated active-power waveforms of IBR1--IBR4. (b)~Mode shape of the IBRs' 18--20 Hz mode.} \label{fig:P_and_mode_shape_simu_v2}
\end{figure}

With the EMT model in place, we validate its accuracy by comparing the simulation results and DFR recordings following the plant A trip at~$t\!=\!0~\mathrm{s}$. Taking IBR1 as an example, Fig.~\ref{fig:PMRF_freq_simu}(a) shows that the simulated grid frequency (red trace (i)) is close to the recorded one (blue trace (ii)). Also, Fig.~\ref{fig:PMRF_freq_simu}(b) shows that the fast Fourier transform (FFT) spectra of the simulated and recorded frequencies are similar. We plot the initial-stage simulated active-power responses of IBR1--IBR4 in Fig.~\ref{fig:P_and_mode_shape_simu_v2}(a) and visualize their 18--20 Hz mode shapes in~Fig.~\ref{fig:P_and_mode_shape_simu_v2}(b). Through visual inspection of Fig.~\ref{fig:P_and_mode_shape_simu_v2}, we find that IBR1 and IBR2 oscillate against IBR3 and IBR4 and have larger oscillation magnitudes after~$t=0~\mathrm{s}$. This observation aligns well with the oscillation recordings in Fig.~\ref{fig:P_and_mode_shape_dfr_v2} and thus validates the accuracy of our EMT model. Note that because not all vendor-provided validated models of IBR1--IBR4 are available, our EMT model aims to capture the oscillation mechanism instead of recreating the exact same recorded dynamics shown in Figs.~\ref{fig:PMRF_Freq} and~\ref{fig:P_and_mode_shape_dfr_v2}.

\subsubsection{Mitigation Methods} \label{subsec:methods}
In a last step, we propose \textcolor{black}{two} mitigation methods for the 18--20 Hz oscillations. Method~1 is to revise the IBR1 and IBR2 inverter-level P/f droop constant from~$3.0\%$ to~$4.0\%$, and Method~2 is to reduce their phase-locked loop (PLL) proportional gains from 0.15 to 0.10\footnote{\color{black}In Fig.~\ref{fig:PMRF_Freq}, IBR1 and IBR2 adopted large PLL proportional gains of 0.15 to satisfy KIUC's fast frequency response requirements, i.e., reaction time $\leq 50~\mathrm{ms}$. But the faster IBR response speed  reduces the damping of the 18--20 Hz system mode in the studied oscillation event.}.  To achieve this, we performed a sensitivity study for \url{~40} SG or IBR controller parameters using the EMT model. In this study, we found that perturbing the inverter-level P/f droop constants and PLL parameters of IBR1 and IBR2 have larger impacts on the simulated oscillations. Based on this finding and some additional trial-and-error efforts, we propose the \textcolor{black}{two} mitigation methods for the 18--20 Hz oscillations. \textcolor{black}{(KIUC has since converted IBR1 from GFL to grid-forming mode; this has been observed to effectively mitigate the oscillations.)}

\section{Simulation Validation} 
This section validates the effectiveness of our \textcolor{black}{two} oscillation mitigation methods in Section~\ref{subsec:methods}. This validation is performed by comparing the EMT simulation results adopting method~1 or 2 against the base case results with no mitigation method, i.e., trace~(i) in Fig.~\ref{fig:PMRF_freq_simu}.

\subsubsection{Validation of Mitigation Method~1} \label{ssec:veri_m1} 
\begin{figure}[t!]
  \centering
  \includegraphics[width=1\linewidth]{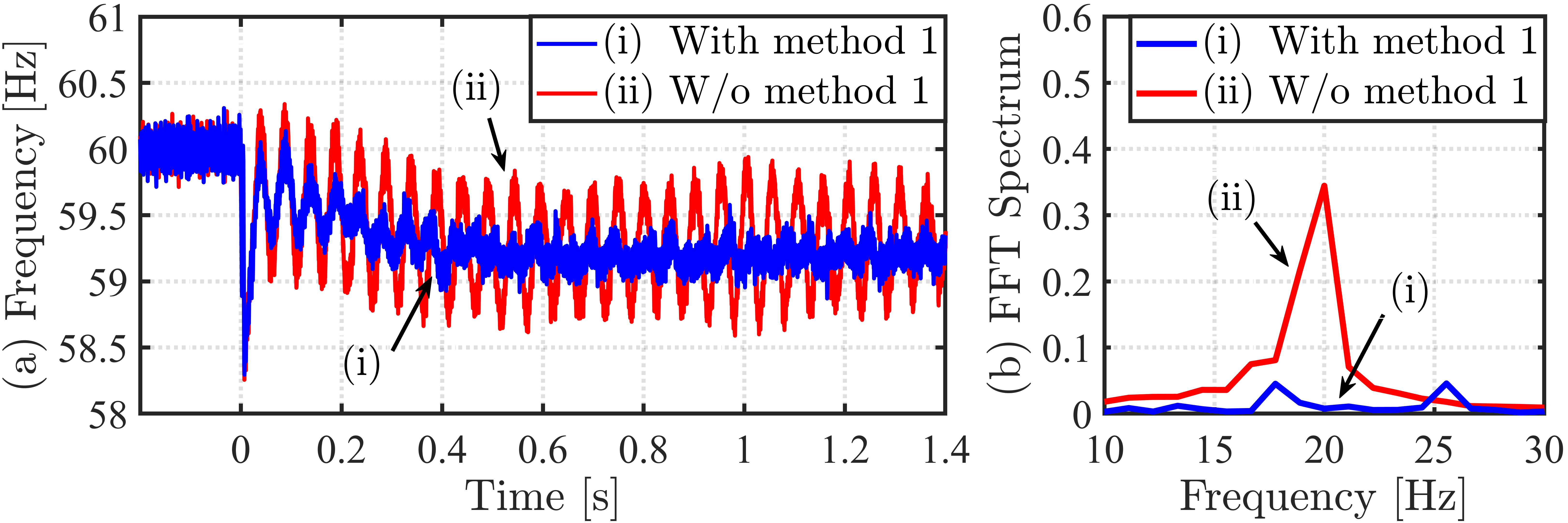}
  \caption{Validation of our Method 1, which aims to mitigate the 18--20 Hz oscillations. (a)~Simulated grid frequencies measured at IBR1 with and without Method~1. (b)~FFT analysis results of simulated grid frequencies.} \label{fig:PMRF_freq_simu_m1}
\end{figure}

In Fig.~\ref{fig:PMRF_freq_simu_m1}(a), we plot the simulated frequency after adopting Method~1 (blue trace~(i)) and the base case (red trace~(ii)). It can be found that the blue trace (i) has a much smaller oscillation magnitude after~$t=0~\mathrm{s}$. Also, the FFT results in Fig.~\ref{fig:PMRF_freq_simu_m2} show that adopting Method~1 (blue trace~(i)) removes the 18--20 Hz oscillation peak that exists in the base case (red trace~(ii)).

\subsubsection{Validation of Mitigation Method~2} 
\begin{figure}[t!]
  \centering
  \includegraphics[width=1\linewidth]{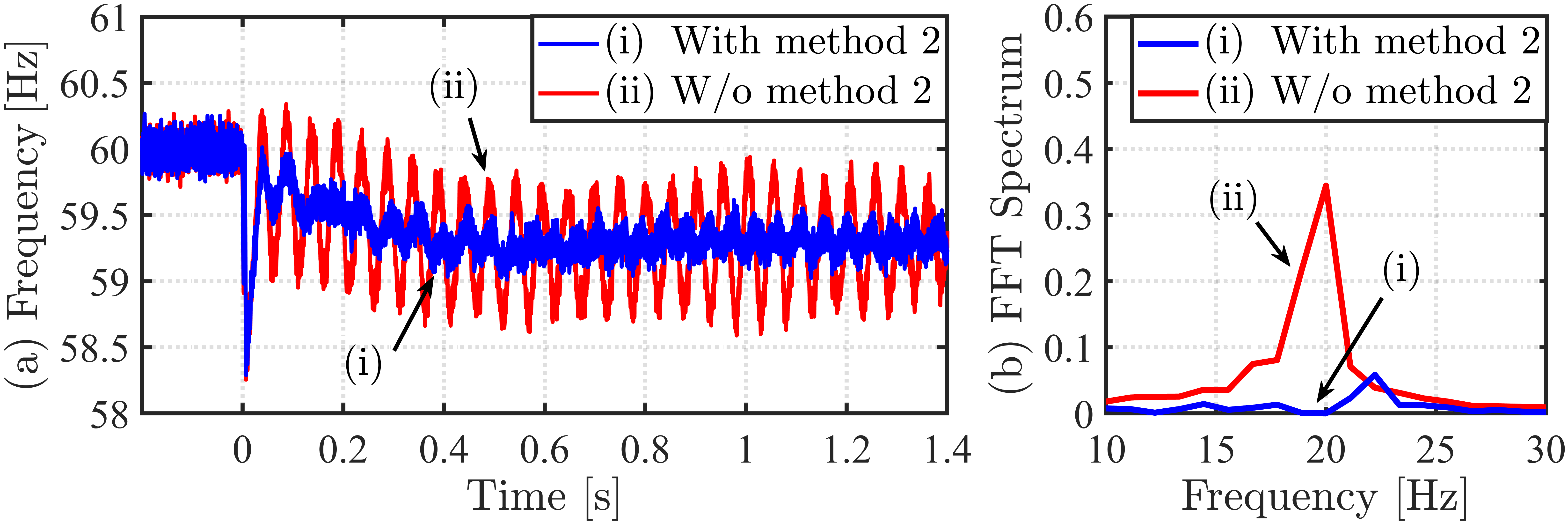}
  \caption{Validation of our Method 2, which aims to mitigate the 18--20 Hz oscillations. (a)~Simulated grid frequencies measured at IBR1 with and without Method~2. (b)~FFT analysis results of simulated grid frequencies.} \label{fig:PMRF_freq_simu_m2}
\end{figure}

We repeat the validation process in~Section~\ref{ssec:veri_m1} for Method~2. As shown in Figs.~\ref{fig:PMRF_freq_simu_m2}(a) and (b), adopting Method~2 also allows us to remove the 18--20 Hz oscillations in the base case. So, both methods~1 and~2 can effectively remove the 18--20 Hz oscillation.

\section{Concluding Remarks} 
In this letter, we report the 18--20 Hz oscillations observed at 05:30 am HST on November 21, 2021. We also analyze the oscillation event with both measurement- and model-based methods. Through our analysis, we identify the oscillation sources, recreate the oscillation with EMT simulations, and propose \textcolor{black}{two} effective mitigation methods. \textcolor{black}{The mitigation methods included both a revision to the IBR P/f droop constant and changes to the proportional gains of the PLL.} Compelling directions for future work include performing detailed small-signal analysis, further identifying the root cause of the oscillations, using grid-forming IBR controls as oscillation mitigation methods, and validating the mitigation methods with power hardware-in-the-loop tests or field experiments.

\section*{Acknowledgment}
This work was authored in part by the National Renewable Energy Laboratory, operated by Alliance for Sustainable Energy, LLC, for the U.S. Department of Energy (DOE) under Contract No. DE-AC36-08GO28308. This material is based upon work supported by the U.S. Department of Energy's Office of Energy Efficiency and Renewable Energy (EERE) under the Solar Energy Technologies Office Award Number 37772. The U.S. Government retains and the publisher, by accepting the article for publication, acknowledges that the U.S. Government retains a nonexclusive, paid-up, irrevocable, worldwide license to publish or reproduce the published form of this work, or allow others to do so, for U.S. Government purposes. The views expressed herein do not necessarily represent the views of the U.S. Department of Energy or the United States Government.

\bibliographystyle{IEEEtran}
\nocite{*}
\bibliography{paper}
%\balance

\end{document}